\documentclass[11pt]{article}
\usepackage{amsmath,amssymb,amsfonts,amstext} \usepackage{hyperref}
\begin{document}

\title{\hfill{\bf \small OUTP-08-05P}\\
       Fine tuning and the ratio of tensor to scalar density fluctuations
       from cosmological inflation}
\author{Shaun Hotchkiss, 
        Gabriel Germ\'an,\thanks{On sabbatical leave from 
Instituto de Ciencias F\'isicas, Universidad Nacional Aut\'onoma de M\'exico}~
        Graham G Ross\thanks{Corresponding author: 
\texttt{g.ross@physics.ox.ac.uk}}~
        and Subir Sarkar \\
\normalsize\textit{Rudolf Peierls Centre for Theoretical Physics,} \\
\normalsize\textit{University of Oxford, 1 Keble Road, Oxford, OX1 3NP, UK}}
\date{}
\maketitle
\begin{abstract}
  The form of the inflationary potential is severely restricted if one
  requires that it be natural in the technical sense, i.e. terms of
  unrelated origin are not required to be correlated. We determine the
  constraints on observables that are implied in such natural
  inflationary models, in particular on $r$, the ratio of tensor to
  scalar perturbations. We find that the naturalness constraint does
  not require $r$ to be lare enough to be detectable by the
  forthcoming searches for B-mode polarisation in CMB maps. We show
  also that the value of $r$ is a sensitive discriminator between
  inflationary models.
\end{abstract}

The nature of the density perturbations originating in the early
universe has been of great interest both observationally and
theoretically. The hypothesis that they were generated during an early
period of inflationary expansion has been shown to be consistent with
all present observations. The most discussed mechanism for inflation
is the `slow roll' of a weakly coupled `inflaton' field down its
potential --- the near-constant vacuum energy of the system during the
slow-roll evolution drives a period of exponentially fast expansion
and the density perturbations have their origin as quantum
fluctuations in the inflaton energy density.

In such models the detailed structure of the density perturbations
which give rise to the large scale structure of the universe observed
today depends on the nature of the inflationary potential in the field
region where they were generated. Boyle, Steinhardt and Turok
\cite{Boyle:2005ug} have argued that ``naturalness'' imposes such
strong restrictions on the inflationary potential that one may derive
interesting constraints on observables today. They concluded that in
theories which are ``natural'' according to their criterion, the
spectral index of the scalar density perturbations is bounded as
$n_\mathrm{s}<0.98$, and that the ratio of tensor-to-scalar
perturbations satisfies $r>0.01$ provided $n_\mathrm{s}>0.95$, in
accord with then current measurements \cite{Spergel:2003cb}. Such a
lower limit on the amplitude of gravitational waves is of enormous
interest as there is then a realistic possibility of detecting them as
`B-mode' polarisation in CMB sky maps (see
e.g. \cite{Efstathiou:2007gz}) and thus verifying a key prediction of
inflation.

Of course these conclusions are crucially dependent on the definition
of naturalness.  In this paper we re-examine this important issue and
argue that the criterion proposed by Boyle {\em et al} does not
capture the essential aspects of a {\em physically} natural theory. We
propose an alternative criterion that correctly reflects the
constraints coming from underlying symmetries of the theory and we use
this to determine a new bound on $r$ that turns out to quite opposite
to the previously inferred one. We emphasise that our result, although
superficially similar to the `Lyth bound' \cite{Lyth:1996im}, follows
in fact from different considerations and in particular makes no
reference to how long inflation lasts.

Inflation predicts a near scale-invariant spectrum for the scalar and
tensor fluctuations, the former being in reasonable agreement with
current observations. Here we explore the predictions for {\em
  natural} models involving a single inflaton at the time the density
perturbations are produced. Models with two or more scalar fields
affecting the density perturbations require some measure of fine
tuning to relate their contribution to the energy density, whereas the
single field models avoid this unnatural aspect. In order to
characterize the inflationary possibilities in a model independent way
it is convenient to expand the inflationary potential about the value
of the field $\phi_\mathrm{H}$ just at the start of the observable
inflation era, $\sim 60$ e-folds before the end of inflation when the
scalar density perturbation on the scale of our present Hubble
radius~\footnote{ Numerically this is $H_0^{-1} \simeq
  3000h^{-1}$~Mpc, where $h \equiv H_0/100$~km~s$^{-1}$~Mpc$^{-1} \sim
  0.7$ is the Hubble parameter. The density perturbation is measured
  down to $\sim 1$~Mpc, a spatial range corresponding to $\sim 8$
  e-folds of inflation.} was generated, and expand in the field
$\phi^\ast \equiv \phi - \phi_\mathrm{H}$ \cite{German:2001tz}. Since
the potential must be very flat to drive inflation, $\phi ^{\ast }$
will necessarily be \emph{small} while the observable density
perturbations are produced, so the Taylor expansion of the potential
will be dominated by low powers of $\phi^\ast$:
\begin{equation}
V (\phi^\ast) = V(0) + V^\prime(0) \phi^\ast 
               + \frac{1}{2}V^{\prime\prime}(0)\phi^{\ast 2} + \ldots 
\label{expand}
\end{equation}
The first term $V(0)$ provides the near-constant vacuum energy driving
inflation while the $\phi^\ast$-dependent terms are ultimately
responsible for ending inflation, driving $\phi^\ast$ large until
higher-order terms violate the slow-roll conditions. These terms also
determine the nature of the density perturbations produced, in
particular the departure from a scale-invariant spectrum.

The observable features of the primordial density fluctuations can
readily be expressed in terms of the coefficients of the Taylor series
\cite{German:2001tz}. It is customary to use these coefficients first
to define the slow-roll parameters $\epsilon$ and $\eta$
\cite{Liddle:2000cg} which must be small during inflation:
\begin{equation}
\epsilon \equiv \frac{M^2}{2}\left(\frac{V^\prime(0)}{V(0)}\right)^2 \ll 1,
\qquad 
|\eta| \equiv M^2 \left\vert 
\frac{V^{\prime\prime}(0)}{V(0)}\right\vert \ll 1,  
\label{slowroll}
\end{equation}
where $M$ is the reduced Planck scale, $M=2.44\times 10^{18}$~GeV. In
terms of these the spectral index is given by
\begin{equation}
n_\mathrm{s} = 1 + 2\eta - 6\epsilon, 
\label{spectral}
\end{equation}
the tensor-to-scalar ratio is
\begin{equation}
r = 16\epsilon, 
\label{indicetensorial}
\end{equation}
and the density perturbation at wave number $k$ is 
\begin{equation}
\delta_\mathrm{H}^2 (k) = \frac{1}{150\pi^2}\frac{V(0)}{\epsilon M^4} .
\label{densitypert}
\end{equation}
Finally the `running' of the spectral index is given by
\begin{equation}
n_\mathrm{r} \equiv \frac{\mathrm{d}n_\mathrm{s}}{\mathrm{d}\ln k} 
             = 16\epsilon\eta - 24\epsilon^2 - 2\xi , 
\label{spectraltilt}
\end{equation}
where
\begin{equation}
\xi \equiv M^4 \frac{V^\prime V^{\prime\prime\prime}}{V^2}.
\label{xi}
\end{equation}

At this stage we have four observables, $n_\mathrm{s},$
$n_\mathrm{r}$, $\delta_\mathrm{H}$ and $r$ and four unknown
parameters $V(0)$, $V^\prime(0)$, $V^{\prime\prime}(0)$ and
$V^{\prime\prime\prime}(0)$ which, for an arbitrary inflation
potential, are independent. However for natural potentials these
parameters are related, leading to corresponding relations between the
observables. Observational confirmation of such relations would
provide evidence for the underlying potential, hence crucial clues to
the physics behind inflation.

As discussed above we are considering the class of natural models in
which a single inflaton field dominates when the density perturbations
relevant to the large-scale structure of the universe today are being
produced.\footnote{This does not exclude `hybrid inflation' models in
  which additional fields play a role at the end of inflation.} In
classifying ``natural'' inflation, Boyle {\em et al} imposed a set of
five conditions \cite{Boyle:2005ug}:

\begin{enumerate}

\item The energy density (scalar) perturbations generated by inflation
  must have amplitude $\sim 10^{-5}$ on the scales that left the
  horizon $\approx 60$ e-folds before the end of inflation;

\item The universe undergoes at least $N > 60$ e-folds of inflation;

\item After inflation, the field must evolve smoothly to an analytic
  minimum with $V = 0$;

\item If the minimum is metastable, then it must be long-lived and $V$
  must be bounded from below;

\item Inflation must halt and the universe must reheat without
  spoiling its large-scale homogeneity and isotropy.

\end{enumerate}
They proposed that the level of fine-tuning for potentials satisfying
the above conditions should be measured by the integers $Z_{\epsilon,
  \eta}$ that measure the number of zeros that $\epsilon$ and $\eta$
and their derivatives undergo within the last 60 e-folds of inflation
\cite{Boyle:2005ug}. Here we argue that such a measure does {\em not}
capture the essential character of physical naturalness.

At a purely calculational level this is illustrated by the fact that
it is necessary to impose an (arbitrary) cut-off on the number of
derivatives included in the criterion.\footnote{Boyle {\em et al}
  imposed the cut-off at 15 derivatives (L. Boyle, private
  communication).} This is necessary because $\epsilon$ and $\eta$ are
defined in terms of the ratio of first or second order derivatives of
the potential to the potential itself, so {\em all} higher order
derivatives must be considered separately when counting the total
number of zeros. The difficulty follows from the observation that, as
far as naturalness is concerned, it is the inflaton potential that is
the primary object, being restricted by the underlying symmetries of
the (effective) field theory describing the inflaton dynamics. As
stressed by 't Hooft \cite{'t Hooft:1979bh}, a {\em natural} theory is
one in which all terms in the Lagrangian allowed by the underlying
symmetries of the theory are present, with no relations assumed
between terms unrelated by the symmetries.

It is important however to note that such natural potentials do {\em
  not} preclude significant contributions from unrelated terms. Indeed
such contributions are inevitable if, for example, the inflation field
is moving from small to large field values. For small field values the
lowest allowed power in the inflaton field is likely to be the most
important, but at larger field values higher powers will ultimately
dominate. For this reason the last four conditions have a different
character to the first in that they involve the end of inflation when
naturalness does {\em not} require that a single term in the inflation
potential should dominate. For example in inflationary models with the
inflaton rolling from small to large field values, the higher powers
can cause the potential to evolve smoothly to an analytic minimum with
$V=0$ or govern the properties of an unstable minimum. Similarly it
may be these higher powers that cause inflation to halt and the
universe to reheat without affecting the predictions for the
observable density perturbation. Given the freedom there is in
choosing these higher powers (non-renormalisable terms in the
effective field theory description), it is {\em always} possible to
find a model in which the end of inflation is satisfactory without
violating the naturalness constraints \cite{German:2001tz}. On the
other hand the range of $\phi^\ast$ relevant during the production of
density perturbations is quite small (corresponding to only 8 of the
$\sim 60$ e-folds of inflation) and so it is reasonable to suppose
that unrelated terms do not simultaneously contribute significantly to
the generation of the observed density perturbation. Although we have
made this argument in the context of `new inflation' models where the
inflaton field evolves from low to high values, similar considerations
apply to the other natural models.  For the case where the underlying
symmetry is a Goldstone symmetry it is still possible to change the
end of inflation in a natural way through the effect of a second
`hybrid' field. Given these considerations we do not need to impose
the last four conditions when determining the phenomenological
implications of natural inflationary models. However we will comment
on how these conditions can indeed be satisfied for the various
classes of inflation potential.

Our definition of naturalness is the standard one in particle physics
\cite{'t Hooft:1979bh}, viz. pertaining to a potential whose form is
guaranteed by a symmetry. This should apply at the time the observable
density perturbations are being produced.  What form can such natural
potentials take? The relevant symmetries that have been identified
capable of restricting the scalar inflaton potential are relatively
limited. The most direct are Abelian or non-Abelian symmetries, either
global or gauge, and continuous or discrete. For a single field
inflation model these will either limit the powers of the inflaton
field that may appear in the potential or, if the inflaton is a
pseudo-Goldstone mode, require a specific form for the potential. Less
direct constraints occur in supersymmetric theories where the scalar
inflaton field is related to a fermionic partner. In this case chiral
symmetries of the associated fermion partner and $R$-symmetries may
further restrict the form of the potential.\footnote{In supersymmetric
  theories it is the superpotential that is constrained by the
  underlying symmetries. The resulting scalar potential has natural
  relations between different powers of the inflaton field.} As
observed earlier \cite{Ross:1995dq}, such symmetries are very
promising for eliminating the fine-tuning problem in inflationary
potentials because they can forbid the large quadratic terms in the
inflaton field that, even if absent at tree level, arise in radiative
order in non-supersymmetric theories (unless protected by a Goldstone
symmetry).

We turn now to a discussion of the observable implications of the
natural inflation models. In this we find it useful to classify the
models into two classes, namely those involving small, sub-Planckian
field values only and those that require large, super-Planckian, field
values. Here we use the reduced Planck scale, $M$, to define the sub-
and super- regimes as this is the scale that orders typical higher
order terms in supergravity. In the small-field models we allow for
the possibility of higher order terms which can dominate as the vacuum
expectation value of the inflaton field becomes large. In the large
field models it is necessary to {\em forbid} such higher order terms
since they would otherwise dominate the potential and there should be
an underlying symmetry to enable this to be done.

\bigskip

\section{Small field models}

These potentials are of the `new inflation' form
\begin{equation}
V\left( \phi \right) =\Delta ^{4}\left[1-\lambda \left( \frac{\phi }{\Lambda}
\right)^p\right], 
\label{slowrollpot}
\end{equation}
with a single power of the inflaton field, $\phi$, responsible for the
variation of the potential, plus a constant term driving inflation.
Such a form does not require fine-tuning as the two terms need not be
related and the dominance of a given single power can be guaranteed by
a symmetry \cite{Ross:1995dq}. Since the slow-roll parameters get no
contribution from the constant term their main contribution will
necessarily come from the leading term involving the inflaton field
and the naturalness condition is trivially satisfied because this is
dominated by a single power of $\phi$. From Eq.(\ref{densitypert}) it
is clear that $\delta_\mathrm{H}$ is the only observable that depends
on $\Delta$, so one can fit its observed value but cannot predict it
without a theory for $\Delta$. The slow-roll parameters are given by
\begin{eqnarray}
\eta &=&-\lambda p(p-1) 
\left(\frac{\phi_\mathrm{H}}{\Lambda}\right)^{p-2}
\left(\frac{M}{\Lambda }\right)^2,\;  
\label{eta2} \\
\epsilon &=&\frac{\lambda^2 p^2}{2}
\left(\frac{\phi_\mathrm{H}}{\Lambda}\right)^{2p-2}
\left(\frac{M}{\Lambda}\right)^2 = \eta^2
\frac{1}{2(p-1)^2}\left(\frac{\phi_\mathrm{H}}{M}\right)^2,\;  
\label{e2}
\\
\xi &=&\lambda^2 p^2 (p-1)(p-2)
\left(\frac{\phi_\mathrm{H}}{\Lambda}\right)^{2p-4} 
\left(\frac{M}{\Lambda}\right)^4 = \eta^2 
\frac{\left(p-2\right)}{\left(p-1\right)} .  
\label{xi2}
\end{eqnarray}

Turning to the other observables let us consider first the cases $p
\geq 2$. Note that the naturalness arguments apply only if
$\phi/\Lambda < 1$ and hence $|\eta| > \epsilon$. In this case
$n_\mathrm{s}$ is effectively determined by $\eta$ alone, so the
measurement of $n_\mathrm{s}$ does not impose a lower bound on
$\epsilon$. Thus the expectation is that $r$ will naturally be small
for this class of models \cite{Ross:1995dq}. To quantify this we note
that $\eta \simeq (1-n_\mathrm{s})/2$ hence
\begin{equation}
r = 16\epsilon = \eta^2 \frac{8}{(p-1)^2}
\left(\frac{\phi_\mathrm{H}}{M}\right)^2.
\end{equation}
This implies that any value $0.9 \lesssim n_\mathrm{s} \lesssim 1$ can
be obtained. Imposing the bound $n_\mathrm{s} > 0.95$ following Boyle
{\em et al} \cite{Boyle:2005ug}\footnote{This is slightly more
  restrictive than the recent WMAP 5-year result: $n_\mathrm{s} =
  0.963_{-0.015}^{+0.014}$ \cite{Hinshaw:2008kr}.}  then requires $r <
0.005$. We emphasise that this makes no explicit reference to the
excursion of the field during inflation, as in the `Lyth bound'
\cite{Lyth:1996im}. Note that the precise value for $r$ depends here
on the value of $\phi_\mathrm{H}$ which, as discussed earlier, is
determined by the higher order terms that may be present in the
potential. Specific examples have been constructed
\cite{German:2001tz} showing that $r$ can be much lower than the bound
given above, even as small as $10^{-16}$. These results are
inconsistent with the {\em lower} bound quoted by Boyle {\em et al}
\cite{Boyle:2005ug} and reflect our different physical interpretation
of naturalness. Finally the prediction for $n_\mathrm{r}$ is
\begin{equation}
n_\mathrm{r} \simeq -2\xi \simeq -0.001 \frac{(p-2)}{(p-1)} .
\end{equation}

The case $p=1$ is special since now $\eta$ and $\xi$ both vanish
giving $r=8(1-n_\mathrm{s})/3 < 0.13$ and $n_\mathrm{r} =
-2(1-n_\mathrm{s})^2/3\simeq 10^{-3}$. This is the {\em only} case of
a sub-Planckian model yielding a large tensor amplitude and it has
been argued \cite{Alabidi:2005qi} that this case cannot be realised in
a complete model due to the requirement that the universe should
undergo at least $\sim 50$ e-folds of inflation. The problem is that
for this case $\epsilon = (1 - n_\mathrm{s})/6$ is large, limiting the
number of e-foldings, which is given by
\begin{equation}
N = \frac{1}{M}\int_{\phi_\mathrm{H}}^{\phi_\mathrm{e}}
\frac{1}{\sqrt{2\epsilon}} \mathrm{d}\phi ,
\end{equation}
where $\phi_\mathrm{e}$ is the field value at the end of inflation.
For sub-Planckian models $\phi_\mathrm{e} \leq M$, hence $N <
1/\sqrt{2\epsilon} = \sqrt{3/(1 - n_\mathrm{s})}$. For the case
$n_\mathrm{s}=0.95$, which gives the large $r$ value, we have only $N
< 8$ e-folds. In this case the effect of higher order terms near the
Planck scale does not help as the linear term already contributes too
much to the slope of the potential and thus limits the number of
e-folds of inflation. The only way out of this is that there should be
a subsequent inflationary era which generates $\sim 40-50$ additional
e-folds of inflation after the $\phi$ field has settled into its
minimum. At first sight this looks like an unnatural requirement.
However we have shown elsewhere \cite{Adams:1997de} that in
supergravity models it is natural to expect some $\sim
3\ln(M/\Lambda)$ e-folds of `multiple inflation' due to intermediate
scale symmetry breaking along `flat directions', where $\Lambda^4$ is
the magnitude of the potential driving this subsequent period of
inflation. Taking $\Lambda \sim 10^{11}$~GeV (typical of the
supersymmetry breaking scale in supergravity models) one generates
$\sim50$ e-folds of inflation.  Although this two-stage inflationary
model appears complicated, it is still natural in the sense discussed
above and should not be ignored as a possibility.

\section{Large field models}

\subsection{Chaotic inflation}

The simplest potential involves a single power of the inflaton

\begin{equation}
V (\phi) = \lambda\frac{\phi^p}{\Lambda^{p-4}}\ ,  
\label{chaotic}
\end{equation}
where we have allowed for the possibility that the scale, $\Lambda$,
relevant for higher dimensional terms in the effective potential need
not be the Planck scale but can correspond to the mass of some heavy
states that have been integrated out in forming the effective
potential. Expanding around $\phi = \phi_\mathrm{H}$ yields
\begin{eqnarray}
\epsilon &=& \frac{p^2}{2} \left(\frac{M}{\phi_\mathrm{H}}\right)^2 , \\
\eta &=& p(p-1) \left(\frac{M}{\phi_\mathrm{H}}\right)^2 , \\
\xi &=& p^2 (p-1)(p-2) \left(\frac{M}{\phi_\mathrm{H}}\right)^4 .
\label{slowrollparams}
\end{eqnarray}
The slow-roll conditions, $\epsilon, |\eta| \ll 1$, requires
$M/\phi_\mathrm{H} \ll 1$ which means that inflation occurs for $\phi$
above the Planck scale --- usually called `chaotic inflation'
\cite{Linde:1983gd}.\footnote{In fact ``chaotic'' actually refers to
  the initial conditions for inflation and `chaotic inflation' can
  also be realised in a small-field model \cite{Linde:1984cd}.} In
this case, in order to explain why ever higher order terms $\phi^m,\;m
\rightarrow \infty$, do not dominate, it is necessary to have a
symmetry which forbids such terms. One such (Goldstone) symmetry has
been invoked in a supergravity context \cite{Kawasaki:2000yn},
although it is not known if this can arise in realistic
models. Another recent proposal for a large field potential exploits
monodromy in a D-brane setup but contains no Standard Model sector
which would give rise to large corrections to the slow-roll parameters
\cite{Silverstein:2008sg}. Thus whether such models can actually be realised
remains an open question.

What are the observable implications of this potential? As before,
$\delta_\mathrm{H}$ is the only observable that depends on $\lambda$
so one can fit its observed value but lacking a theory for $\lambda$
this is not a prediction. The other 3 observables are determined in
terms of the parameter $x = M/\phi_\mathrm{H}$ and the power $p$.
\begin{eqnarray}
n_\mathrm{s} &=& 1-x^2 p(p+2) , \\
r &=& 8 p^2 x^2, \\
n_\mathrm{r} &=& -2x^4 p^2 (p+2) .
\end{eqnarray}
From this one sees that the ratio of tensor to scalar fluctuations and
the running of the spectral index are tightly constrained by the
measurement of $n_\mathrm{s}$
\begin{eqnarray}
r &=& \frac{8p}{(p+2)}(1 - n_\mathrm{s}) , \\
n_\mathrm{r} &=& -\frac{2}{(p + 2)}(1 - n_\mathrm{s})^2.
\end{eqnarray}
Note that these results are independent of $\phi_\mathrm{H}$ and so,
as anticipated above, do not depend on exactly when inflation ends.
For the quartic potential $p=4,$ $r=0.27$ and $n_\mathrm{r} \simeq - 8
\times 10^{-4}$. The maximum value of $r$ is $8(1 - n_\mathrm{s})
\simeq 0.4$ with $n_\mathrm{r}=0.$

\subsection{Natural inflation} 

Another class of non-fine-tuned models is based on an approximate
Goldstone symmetry \cite{Freese:1990rb}, often called `natural
inflation' (although it should now be clear that this is not the {\em
  only} natural possibility). In this case the potential is not a
simple polynomial but has the form
\begin{equation}
V(\phi) = \Delta^4 \left(1 + \cos\frac{\phi}{f}\right).  
\label{freese}
\end{equation}
The slow-roll parameters are:
\begin{eqnarray}
\epsilon &=& \frac{1}{2}\left( \frac{M}{f}\right)^2 
\frac{\left(\sin\frac{\phi_\mathrm{H}}{f}\right)^2}
{\left(1 + \cos\frac{\phi_\mathrm{H}}{f}\right)^2} , \\
\eta &=& 
-\left(\frac{M}{f}\right)^2 
\frac{\cos\frac{\phi_\mathrm{H}}{f}}
{\left(1+\cos\frac{\phi_\mathrm{H}}{f}\right)} , \\
\xi &=& -\left(\frac{M}{f}\right)^4 
\frac{\left(\sin\frac{\phi_\mathrm{H}}{f}\right)^{2}}
{\left(1 + \cos\frac{\phi_\mathrm{H}}{f}\right)^2} .
\end{eqnarray}
For these to be small we require $f > M$. Unlike
the previous case the predictions now depend sensitively on
$\phi_\mathrm{H}$ and hence on the related value of the field at the
{\em end} of inflation.

If the end of inflation is determined, as has usually been {\em
  assumed}, by the steepening of the above potential then
$\phi_\mathrm{H}$ has a value such that $\epsilon$ and $\eta$ are
comparable. In this case $\epsilon$ can be close to its slow-roll
limit, particularly interesting for tensor fluctuations which can now
be large. Imposing the bound $n_\mathrm{s} > 0.95$
\cite{Hinshaw:2008kr} implies $0.02 < r <0.2$ $\cite{Freese:2008if}$,
the range corresponding to the variation of $\phi_\mathrm{H}$ with $f$
for allowed values of $f$. As with the other models, the running is
small, $n_\mathrm{r} \sim {\cal O}(10^{-3})$.

However it may be more natural for inflation to end much earlier due to a
second (hybrid) field. Then $\phi_\mathrm{H}$ is reduced so
that $\sin(\phi_\mathrm{H}/f)$ can be small, hence $|\eta| \gg
\epsilon$. In this case $r$ will be (arbitrarily) small, being
proportional to $\epsilon$, The running is also very small,
$n_\mathrm{r} \simeq 12\epsilon\eta$.

\section{Conclusions}

We have discussed natural possibilities for the inflationary
potential. From this it is clear that the gravitational wave signal
need not be large enough to be observable as argued by Boyle {\em et
  al} \cite{Boyle:2005ug}.

The models considered fall broadly into two classes. The first has
$\epsilon$ comparable in magnitude to $\eta$ hence $r$ can saturate
the upper bound of 0.4 implied by the slow-roll constraint. A
characteristic of these models however is that inflation occurs only
at field values higher than the Planck scale and it is not clear if
this can be naturally realised. An interesting exception is the `new
inflation' model with a leading {\em linear} term in the inflaton
field which however requires a subsequent period of inflation to
create our present Hubble volume.

The second class of models has $\eta$ larger than $\epsilon$. These
are indeed natural but there is no lower bound to $r$ and the upper
bound is (unobservably) small. A characteristic of most such models is
that inflation occurs at low field values, much below the Planck
scale. Examples of this are provided by `new inflation' where $r$ is
bounded from {\em above} by 0.005 and is usually much below this
bound.  A large-feld exception to this is a modified form of `natural
inflation' where a hybrid field ends inflation early.

To summarise, in models that are not fine-tuned, the amplitude of
density perturbations and the spectral index are not predicted, being
determined by free parameters of the model. However the
tensor-to-scalar ratio, $r$, and the running, $n_\mathrm{r}$, are
determined in terms of the spectral index. The ratio $r$ provides a
sensitive discriminator of the natural models but there is no
requirement that it be greater than 0.01 even if the spectral index is
bounded from below $n_\mathrm{s}>0.95$.\footnote{ This conclusion is
  in agreement with another analysis of specific models
  \cite{kinney:2006gm}.} All the natural models have the running
small, $n_\mathrm{r} \sim (1 - n_\mathrm{s})^2$, so observation of a
much larger value would indicate a departure from naturalness, perhaps
because more than one inflaton field is active at the time density
perturbations are generated. This in turn would suggest there should
be a departure from a near-Gaussian distribution of the perturbations.

\section{Acknowledgements}

G.G. acknowledges support from DGAPA, UNAM and the hospitality of the
Rudolf Peierls Centre, Oxford. S.S. acknowledges a STFC Senior
Fellowship (PP/C506205/1) and the EU network `UniverseNet'
(MRTN-CT-2006-035863). We thank Latham Boyle for helpful
correspondance.

\end{document}